# Casimir effect and Lorentz invariance violation

S. A. Alavi

*Department of physics, Hakim Sabzevari university, P. O. Box 397, Sabzevar, Iran.*

*E-mail:* s.alavi@hsu.ac.ir

The Casimir effect is one of the most direct manifestations of the existence of the vacuum quantum fluctuations, discovered by H. B Casimir in 1948. On the other hand, Lorentz invariance is one of the main and basic concepts in special relativity, which states that, the laws of physics are invariant under Lorentz transformation. In this work, we calculate the corrections imposed by LIV on Casimir effect (force). This may provide a direct probe to test LIV in nature.

**Introduction**

Symmetries play a fundamental role in theoretical physics. Lorentz and CPT symmetries are fundamental in the Standard Model. Lorentz invariance is also in the foundation of the general relativity theory. General relativity and the standard-model of particle physics provide a very successful framework for describing and explaining most of the observed physical processes and phenomena in nature. In recent years there has been considerable scientific interest in the possibility of Lorentz invariance violation (LIV). It would be a major discovery and could potentially provide valuable information about a possible theory of quantum gravity. The general framework characterizing such violations is the Standard-Model Extension (SME), which incorporates General Relativity and the Standard Model. This violation can occur for photons, neutrinos, muon spin-precession, atoms as well as hadrons such as kaons, protons and neutrons. Extensive experimental efforts have been made to set up upper bounds on the LIV coefficients in all sectors of the SME.

**Casimir effect and Lorentz invariance violation**

The Lagrangian density for the photon sector of the minimal SME is as follows [1]:

$$L_{photon} = -\frac{1}{4}F_{\mu\nu}F^{\mu\nu} - \frac{1}{4}(K_F)_{k\lambda\mu\nu}F^{k\lambda}F^{\mu\nu} + \frac{1}{2}(K_{AF})^k \epsilon_{k\lambda\mu\nu}A^\lambda F^{\mu\nu} - j^\mu A_\mu \quad (1)$$

The small coefficients $(k_{AF})^\kappa, (k_F)_{\kappa\lambda\mu\nu}$ control the Lorentz invariance violation (LIV) which are considered constant. Variation of this Lagrangian yields the inhomogeneous equations of motion (modified Maxwell equations) [1]:

$$\partial_\alpha F_\mu^\alpha + (K_F)_{\mu\alpha\beta\gamma}\partial^\alpha F^{\beta\gamma} + (K_{AF})^\alpha \epsilon_{\mu\alpha\beta\gamma}F^{\beta\gamma} + j_\mu = 0 \quad (2)$$

To study the effects of LIV on electromagnetic phenomena, a new sets of coefficients are introduced [1]:

$$\left(\kappa_{DE}\right)^{jk} = -2(k_F)^{0j0k}$$
$$\left(\kappa_{HB}\right)^{jk} = \frac{1}{2}\epsilon^{jpq}\epsilon^{krs}(k_F)^{pqrs} \quad (3)$$
$$\left(\kappa_{DB}\right)^{jk} = -\left(\kappa_{HE}\right)^{kj} = \left(k_F\right)^{0jpq}\epsilon^{kpq}$$

The electric field $\vec{E}$ ($\vec{D}$) and the magnetic field $\vec{B}$ ($\vec{H}$), which satisfy in the relations $\vec{D} = \epsilon\vec{E}$ and $\vec{H} = \frac{1}{\mu}\vec{B}$ also receive corrections due to LIV as follows:

$$\vec{D} = (\epsilon + \kappa_{DE}^{vaccum})\vec{E} + \kappa_{DB}^{vaccum}\vec{B}$$
$$\vec{H} = \left(\frac{1}{\mu} + \kappa_{HB}^{vaccum}\right)\vec{B} + \kappa_{HE}^{vaccum}\vec{E} \quad (4)$$

The electromagnetic energy density is then given by the following expression:

$$u_{LIV} = \frac{1}{2}[(\epsilon + \kappa_{DE}^{vaccum})\vec{E} + \kappa_{DB}^{vaccum}\vec{B}].\vec{E} + \frac{1}{2}\left[\left(\frac{1}{\mu} + \kappa_{HB}^{vaccum}\right)\vec{B} + \kappa_{HE}^{vaccum}\vec{E}\right].\vec{B} \quad (5)$$

After some calculations and using the fact that $(\kappa_{DB})^{jk} = -(\kappa_{HE})^{kj}$ [1], one can show that: $u_{LIV} = u_{LI} + \Delta u_{LI}$ where $u_{LIV}$, $u_{LI}$ and $\Delta u_{LI}$ are the electromagnetic energy density in thevpresence of Lorentz invariance violation, the Lorentz invariant electromagnetic energy density and the corrections on it, due to LIV, respectively. The energy density could be written as $u_{LIV} = (1+L)u_{LI}$ where:

$$u_{LI} = \frac{1}{2}\left(\varepsilon \vec{E}^2 + \frac{1}{\mu}\vec{B}^2\right), \Delta u_{LI} = \frac{1}{2}\left(\kappa_{DE}^{vaccum}\vec{E}^2 + \kappa_{HB}^{vaccum}\vec{B}^2\right) = L u_{LI},$$

$$L = \frac{\left(\kappa_{DE}^{vaccum}\vec{E}^2 + \kappa_{HB}^{vaccum}\vec{B}^2\right)}{\left(\varepsilon \vec{E}^2 + \frac{1}{\mu}\vec{B}^2\right)}$$

(6)

The quantized Hamiltonian can be written as:

$$H = \frac{1}{2}\sum_k \omega_k (a_k^\dagger a_k + a_k a_k^\dagger) = \sum_k \omega_k (a_k^\dagger a_k + \frac{1}{2}) = \sum_k \omega_k (n_k + \frac{1}{2})$$

(7)

where $\omega_k = \omega_k^0 + \Delta\omega_k^0$. $\omega_k^0$ is the frequency in the absence of LIV and $\Delta\omega_k^0$ is the correction due to LIV. $\omega_k$ may be written as $\omega_k = (1+L)\omega_k^0$ where $L \to 0$ in the absence of LIV.

Now we obtain the Casimir effect for two perfectly conducting metals. One can describe electromagnetic field as if it is made of two scalar massless fields, one satisfying Dirichlet boundary conditions and the other subject to Neumann conditions[2]:

$$\Box^2 \phi = 0$$

$$\phi(x,t) \sim \sin(\frac{\pi n x}{a}) e^{i\vec{k}\cdot\vec{x}} e^{-i\omega_{kn}^D t}$$

$$\omega_{kn}^D = \left[\left(\frac{\pi n}{a}\right)^2 + k^2\right]^{\frac{1}{2}}, n = 1,2,3...$$

(8)

$$\phi(x,t) \sim \cos(\frac{\pi n x}{a}) e^{i\vec{k}\cdot\vec{x}} e^{-i\omega_{kn}^N t}$$

$$\omega_{kn}^N = \left[\left(\frac{\pi n}{a}\right)^2 + k^2\right]^{\frac{1}{2}}, n = 0,1,2,3...$$

So the sum of all the eigenvalues is [2]:

$$\varepsilon_0(a,L) = \frac{1}{2}(1+L)\left(\frac{1}{a}\sum_{n=1}^{\infty}\int\frac{d^2 k_T}{(2\pi)^2}\omega_{kn}^D(a) + \frac{1}{a}\sum_{n=0}^{\infty}\int\frac{d^2 k}{(2\pi)^2}\omega_{kn}^N(a)\right)$$

(9)

$$= \frac{1}{2a}(1+L)\left(\sum_{n=1}^{\infty} + \sum_{n=0}^{\infty}\right)\int\frac{d^2 k_T}{(2\pi)^2}\left[\left(\frac{\pi n}{a}\right)^2 + k_T^2\right]^{\frac{1}{2}}$$

$$= \frac{1}{2a}\left(\int\frac{d^2 k_T}{(2\pi)^2}\left(k_T^2\right)^{\frac{1}{2}} + \frac{1}{a}\sum_{n=1}^{\infty}\int\frac{d^2 k_T}{(2\pi)^2}\left[\left(\frac{\pi n}{a}\right)^2 + k_T^2\right]^{\frac{1}{2}}\right)$$

Using zeta-function regularization method which starts by replacing the power $\frac{1}{2}$ in the integrand by $-\frac{s}{2}$, we have [2]:

$$\varepsilon_0(s;a,L) = \frac{1}{2a}(1+L)\int \frac{d^2k_T}{(2\pi)^2}\left(k_T^2\right)^{\frac{-s}{2}}$$

$$+\frac{1}{a}(1+L)\sum_{n=1}^{\infty}\int \frac{d^2k_T}{(2\pi)^2}\left[\left(\frac{\pi n}{a}\right)^2 + k_T^2\right]^{\frac{-s}{2}}$$

$$\equiv (1+L)\left[\varepsilon_0^{(1)}(s;a,L) + \varepsilon_0^{(2)}(s;a,L)\right]$$

(10)

After some calculations, we will have:

$$\varepsilon_0^2(L,a) = (1+L)\varepsilon_0^2 \tag{11}$$

Following the same method as Ref.[2], we obtain the expression for the pressure in the presence of LIV:

$$P(L,a) = (1+L)P(a) \propto \frac{(1+L)}{a^4} \tag{12}$$

It is observed that, the effects of LIV on Casimir force (pressure) consists in a multiplicative factor.

Using the accuracy of experimental measurement of Casimir Force which is 1.6 pN, one can put an upper bound on the LIV parameter. The Casimir force for two perfectly conducting parallel plates of area 'A' separated by a distance 'a' is given by [3]:

$$F(d) = \frac{-\pi^2 \hbar c}{240}\frac{A}{a^4} \tag{13}$$

It is a strong function of 'a' and is measurable only for $a < 1$ micrometer. If we take $a \approx 0.01$ micometer, the accuracy of the mesurment of the Casimir force $\approx 1\,pN = 10^{-12}N$ [3] and a 1.25 cm diameter optically polished sapphire disk as the plate, we obtain the following bound for the LIV parameter "L":

$$L \leq 1.6 \times 10^{-5} \tag{14}$$

**References:**

[1]. Quentin G. Bailey, Alan Kostelecky, Phys.Rev.D 70 (2004) 076006.
[2]. E. Elizalde, A. Romeo, Am. J. Phys. 59 (1991) 8.
[3]. U. Mohideen , Anushree Roy, Phys. Rev. Lett. 81 (1998) 4549.